\documentclass{ptephy_hep}
\usepackage{amsmath,amssymb}
\usepackage{bm}
\usepackage[dvipdfmx,implicit=false,colorlinks=true,linkcolor=blue]{hyperref}

\newcommand{\eq}{\begin{equation}}
\newcommand{\eqend}{\end{equation}}
\newcommand{\ovl}{\overline}

\begin{document}

\title{The superspace representation of Super Yang--Mills theory on NCG}

\author{\name{\fname{Masafumi} \surname{Shimojo}}{1,\ast}, \name{\fname{Satoshi} \surname{Ishihara}}{2}, 
\name{\fname{Hironobu} \surname{Kataoka}}{2}, \name{\fname{Atsuko} \surname{Matsukawa}}{2}\\ and  \name{
\fname{Hikaru} \surname{Sato}}{2}
}

\address{
\affil{1}{Department of Electronics and Information Engineering, National Institute of Technology, Fukui College, 
Geshicho, Sabae, Fukui 916-8507, Japan}
\affil{2}{Department of Physics, Hyogo University of Education, Shimokume, Kato, Hyogo 673-1494, Japan}
\email{shimo0@ei.fukui-nct.ac.jp} \\
}

\begin{abstract}%
A few years ago, we found the supersymmetric(SUSY) counterpart of the spectral triple which 
specified noncommutative geometry(NCG). Based on "the triple", we considered the SUSY version of 
the spectral action principle and had derived the action of super Yang--Mills theory, minimal 
supersymmetric standard model, and supergravity. In these theories, we used vector 
notation in order to express a chiral or an anti-chiral matter superfield. We also 
represented the NCG algebra and the Dirac operator by matrices which operated on the space of matter field. 
In this paper, we represent the triple in the superspace coordinate system $(x^\mu,\theta,\bar{\theta})$. 
We also introduce "extracting operators" and the new definition of 
the supertrace so that we can also investigate 
the square of the Dirac operator on the Minkowskian manifold in the superspace. 
We finally re-construct the super Yang--Mills theory on NCG in the 
superspace coordinate to which we are familiar to describe SUSY theories.    
\end{abstract}
\subjectindex{B16, B40, B82}
\maketitle
\section{Introduction}
Connes and his co-workers derived the standard model (SM) of high energy physics coupled to 
gravity on the basis on noncommutative geometry(NCG)\cite{connes1,connes2,connes4,connes3}. 
The framework of an NCG is specified by 
a set called a spectral triple$(\mathcal{H}_0,\mathcal{A}_0,\mathcal{D}_0)$\cite{connes0}. 
Here, $\mathcal{H}_0$ is the Hilbert space which consists of the spinorial wave functions of physical matter fields. 
$\mathcal{A}_0$ and $\mathcal{D}_0$ are noncommutative complex algebra and Dirac operator which is a self-adjoint 
operator with compact resolvent. They act on the Hilbert space $\mathcal{H}_0$.    
$Z/2$ grading $\gamma$ and the real structure $\mathcal{J}$ are taken into account to determine the 
KO dimension. 
The Dirac operators has a foliation of 
equivalence classes, the internal fluctuation which is given as follows:
\begin{equation}
\tilde{\mathcal{D}}_0 = \mathcal{D}_0 + A+ JAJ^{-1},\ A=\sum a_i[\mathcal{D},b_i],\ a_i,b_i\in \mathcal{A}.
\end{equation}   
The fluctuation $A+JAJ^{-1}$ for the Dirac operator on the manifold $\mathcal{D}_{0M}
={\rm i}\gamma^\mu\nabla_\mu\otimes 1$ 
gives the gauge vector field, while that for the Dirac operator in the finite space $\mathcal{D}_{0F}$ 
gives the Higgs field\cite{connes7,connes10}.

The action of the NCG model is obtained by the spectral action principle and is expressed by
\begin{equation}
\langle \psi \tilde{\mathcal{D}}_0  \psi \rangle + {\rm Tr}(f(P)). \label{totalaction0}
\end{equation}
Here the first term stands for the matter action and $\psi$ is a fermionic field which belongs to 
$\mathcal{H}_0$.
The second term represents the bosonic part which depends only on the spectrum of the squared Dirac operator
 $P=\tilde{\mathcal{D}}_0^2$ 
and $f(x)$ is an auxiliary smooth function on 
a four-dimensional compact Riemannian manifold without boundary\cite{connes8}. 

The SM has some defects, in particular, has the hierarchy problem. 
It is known that the problem is perfectly remedied by introducing supersymmetry\cite{martin}. 
In order to incorporate the supersymmetry to particle models on concepts of NCG, we have obtained 
"the triple" $(\mathcal{H},\mathcal{A}, \mathcal{D})$ extended from the spectral 
triple on the flat Riemannian manifold and verified its supersymmetry
\cite{paper0,paper1}. Here, $\mathcal{H}$ is the functional space which consists of chiral and 
antichiral supermultiplets that correspond to spinorial and scalar wave functions of $C^\infty(M)$. 
The triple however does not satisfy the axioms of NCG. For an example, 
the commutator $[\mathcal{D},a]$ is not bounded 
for the extended Dirac operator $\mathcal{D}$ and an arbitrary element $a\in \mathcal{A}$, 
because $\mathcal{D}$ includes d'Alembertian which appears in the Klein-Gordon equation. So,   
the triple does not produce a new NCG. When we limit the domain $\mathcal{H}$ to the space of 
the spinorial wave functions $\mathcal{H}_0$, 
the triple reduces to the spectral triple and the whole theory also reduces to the original one. 

We also found the internal fluctuation of the Dirac operator which produced 
vector supermultiplets with gauge degrees of freedom, supersymmetric invariant product of 
elements in $\mathcal{H}$ and the supersymmetric version of the spectral action principle.   
Using these components, we obtained the kinetic and mass terms 
of the matter particle interacted with gauge fields. We also investigated the square of the  
fluctuated Dirac operator and Seeley-DeWitt coefficients of heat kernel expansion, so that we arrived 
at the action of supersymmetric Yang-Mills theory and that of the minimal supersymmetric standard model\cite{paper1,paper2}.

In the above construction of supersymmetric theories on NCG, a chiral or an antichiral superfield, 
an element of the functional space $\mathcal{H}_M$ in the Minkowskian manifold $M$ was described by 
a vector notation such as $(\varphi,\psi^\alpha, F)^T$, where $\varphi,F$ and $\psi^\alpha$ were 
bosonic and spinorial wave functions.  An element of $\mathcal{A}$ and the Dirac operator 
$\mathcal{D}$ were described by matrices which operated on a vector in $\mathcal{H}$. However, in general, 
SUSY theories are formulated in superspace coordinate system 
$(x^\mu, \theta, \bar{\theta})$\cite{WessBagger}. 

In this paper, we will review the supersymmetric Yang-Mills theory on NCG in the ref.\cite{paper1} and 
reconstruct the theory by the superspace representation. 
At first, in Sect.2 and Sect.3, we reconstruct the representations of the basic components, the triple, 
$Z/2$ grading, antilinear operator, supersymmetric invariant products and internal fluctuation of 
the Dirac operator, one by one. 
Secondly, in Sect.4, we reconstruct the supersymmetric version of the spectral action principle. 
In order to represent the square of the fluctuated Dirac operator $\mathcal{D}_M^2$ on $\mathcal{H}_M$ 
in the superspace, we will introduce new operators, which we will call "extracting operators". 
The extracting operators also make it possible to define the supertrace and the representations of 
the other operators, $\mathbb{E}^2$, $\Omega^{\mu\nu}\Omega_{\mu\nu}$, which are necessary to calculate 
coefficients of the heat kernel expansion. Then we will establish the method to 
obtain, by using the superspace coordinate, the action of super Yang-Mills theory on NCG.  
\section{Supersymmetrically extended triple}
In this section, we review the triple,i.e.,the SUSY counterpart of the spectral triple of NCG 
introduced in ref.\cite{paper0,paper1} and 
rewrite the chiral and antichiral superfields which appear in the triple with those 
represented in the superspace coordinates. 

The functional space $\mathcal{H}$ is the product denoted by
\begin{equation}
\mathcal{H}= \mathcal{H}_M \otimes \mathcal{H}_F \label{H}.
\end{equation}
The functional space $\mathcal{H}_M$ on the Minkowskian space-time manifold $M$ 
is the direct sum of two subsets, $\mathcal{H}_+$ and 
$\mathcal{H}_-$:
\begin{equation}
\mathcal{H}_M = \mathcal{H}_+ \oplus \mathcal{H}_- =\{(\Psi_+,0)^T\}+\{(0,\Psi_-)^T\}, \label{HM}
\end{equation}
where $\Psi_+$ is a chiral superfield and $\Psi_-$ 
is an antichiral superfield. 
In our previous paper, these fields are expressed by the vector natation such as $(\varphi_+,\psi_{+\alpha},F_+)$ and 
$(\varphi_-^\ast, \bar{\psi}_-^{\dot{\alpha}},F_-^\ast)$, where $\varphi_\pm$ and $F_\pm$ of $\Psi_\pm$ are complex scalar functions with mass dimension one and 
two, respectively, $\psi_{+ \alpha}$ and $\bar{\psi}_-^{\dot{\alpha}}$ with $\alpha$ and $\dot{\alpha}=1,2$ 
are the Weyl spinors on the space-time $M$ which have mass dimension $\frac{3}{2}$  and transform  
as the $(\frac{1}{2},0)$, $(0,\frac{1}{2})$ representations of the Lorentz group $SL(2,C)$, respectively. 

Now, we represent $\Psi_\pm$ in the superspace as follows: \\
The element $\Psi_+$ in $\mathcal{H}_+$ is given in the superspace coordinate 
$(x_+^\mu=x^\mu+i\theta\sigma^{\mu}\bar{\theta},\theta,\bar{\theta})$ by
\begin{equation}
\Psi_+(x_+) = \varphi_+(x_+)+\sqrt{2}\theta^\alpha\psi_{+\alpha}(x_+)+ \theta\theta F_+(x_+), \label{Psip}
\end{equation}
and the element $\Psi_-$ in $\mathcal{H}_-$ is given in the coordinate  
$(x_-^\mu=x^\mu-i\theta\sigma^\mu\bar{\theta},\theta,\bar{\theta})$ 
by
\begin{equation}
\Psi_-(x_-) = \varphi_-(x_-)^\ast+\sqrt{2}\bar{\theta}_{\dot{\alpha}}\bar{\psi}_-^{\ast\dot{\alpha}}(x_-) \label{Psim}
+\ovl{\theta\theta}F_-^\ast(x_-).
\end{equation}
Hereafter, the argument $(x_\pm)$ of fields and operators denotes the superspace coordinate system 
$(x_\pm^\mu, \theta,\bar{\theta})$
in which those fields and operators are expressed. 

The Z/2 grading $\gamma_M$ of the functional space $\mathcal{H}_M$ is given by 
\begin{equation}
\gamma_M=\left\{
\begin{array}{cc}
-{\rm i}\mathbb{I}_+ & {\rm in\ } \mathcal{H}_+ \\
{\rm i}\mathbb{I}_- & {\rm in \ } \mathcal{H}_-
\end{array},
\right.
\end{equation}
where $\mathbb{I}_+$, $\mathbb{I}_-$ are the identity operators on $\mathcal{H}_+$, $\mathcal{H}_-$ 
which we will describe later. 

The finite space $\mathcal{H}_F$ is the space with the basis of the labels 
$q_L^a$, $q_R^a$, $(q_c^a)_L$ and $(q_c^a)_R$, which correspond to matter particles, antiparticles 
and their superpartners, such as quarks, squarks and auxiliary fields. We express them by the previous notations as follows:
\begin{equation}
Q^a =(q_L^a,q_R^a)^T, \label{Qa}
\end{equation}
for the particle part and
\begin{equation}
Q_c^a =((q_c^a)_L,(q_c^a)_R)^T, \label{Qca}
\end{equation}
for the antiparticle part. Here $a$ is the index, $a=1,\ldots, N$, which denotes  
internal degrees of freedom, $L$ and $R$ denote the eigenstates of 
the $Z/2$ grading $\gamma_F$ for the discrete space, which is defined by
\begin{equation}
\gamma_F(q_L^a)=-1,\ \gamma_F(q_R^a)=1. 
\end{equation}
In order to evade fermion 
doubling \cite{gracia,lizzi}, we impose that the physical 
wave functions obey the following condition:
\begin{equation}
\gamma =\gamma_M\gamma_F ={\rm i}.
\end{equation}
Then for the supermultiplet which is a set of a left-handed 
fermionic matter field and its superpartner and auxiliary field, we have 
\begin{equation}
\Phi_L = q_L^a\otimes (\Psi_+(x_+),0)^T 
 = q_L^a \otimes (\varphi_+ + \sqrt{2}\theta^\alpha \psi_{+\alpha}+\theta\theta F_+,0)^T \label{PsiL}.
\end{equation} 
So the physical wave functions of the chiral supermultiplet amount to
\begin{equation}
\varphi_L  = q_L^a\otimes (\varphi_+,0)^T,\ \  
\psi_{L\alpha}  =   q_L^a\otimes (\psi_{+\alpha},0)^T,\ \  
F_L  = q_L^a\otimes (F_+,0)^T  \label{vphiFRp}.
\end{equation} 
For the physical wave functions of the right-handed fermionic matter field, we have
\begin{equation}
\Phi_R = q_R^a\otimes (0,\Psi_-(x_-))^T = q_R^a\otimes (0, \varphi_-^\ast+\sqrt{2}\bar{\theta}_{\dot{\alpha}}
\bar{\psi}_-^{\dot{\alpha}}+\ovl{\theta\theta}F_-^\ast)^T
\end{equation}
and
\begin{equation}
\varphi_R  = q_R^a \otimes (0,\varphi_-^\ast)^T,\ \ 
\psi_R^{\dot{\alpha}}  = q_R^a \otimes (0,\bar{\psi}_-^{\dot{\alpha}})^T,\ \ 
F_R  = q_R^a \otimes (0,F_-^\ast)^T.  \label{vphiFRm}
\end{equation}

For a state $\Psi \in \mathcal{H}_M$ , its charge conjugate state $\Psi^c$ is given by 
hermitian conjugation $\Psi^\dagger$. The real structure transforms an element 
of $\mathcal{H}_\pm $ into an element of $\mathcal{H}_\mp$, 
automatically,
\begin{align}
\mathcal{J}_M \Psi_+(x_+) = & \Psi_+^\dagger(x_-) =\varphi_+^\ast 
+\sqrt{2}\bar{\theta}\bar{\psi}+\ovl{\theta\theta}F_+^\ast, \label{JM1}\\                        
\mathcal{J}_M \Psi_-(x_-) = & \Psi_-^\dagger(x_+) = \varphi_- +\sqrt{2}\theta\psi_- +\theta\theta F_-, \label{JM2}
\end{align}
and $\mathcal{J}_M$  is commutative with $Z/2$ grading $\gamma_M$.

In the finite space, the antilinear operator $\mathcal{J}_F$ is the 
replacement of the label of particles to those of antiparticles, 
\begin{equation}
\mathcal{J}_F q_L^a = (q_L^a)^c =(q_a^c)_R,\ \  
\mathcal{J}_F q_R^a = (q_R^a)^c=(q_a^c)_L.  \label{JF} 
\end{equation} 
From (\ref{JM1}),(\ref{JM2}),(\ref{JF}), the total real structure $J=\mathcal{J}_M\otimes \mathcal{J}_F$ 
operates on $\mathcal{H}$ as follows:
\begin{align}
J\Phi_L = & \varphi_L^\ast+\sqrt{2}\bar{\theta}
\bar{\psi}_L+\ovl{\theta\theta}F_L^\ast 
=(\varphi^c)_R+\sqrt{2}\bar{\theta}
(\psi^c)_R+\ovl{\theta\theta}(F^c)_R= (\Phi^c)_R, \\
J\Phi_R = & \varphi_R^\ast + \sqrt{2}\theta\bar{\psi}_R+\theta\theta F_R^\ast=
(\varphi^c)_L+\sqrt{2}\theta
(\psi^c)_L+\theta\theta (F^c)_L =(\Phi^c)_L. 
\end{align}

Corresponding to the construction of the functional space (\ref{H})
and (\ref{HM}), 
the algebra $\mathcal{A}$ represented on $\mathcal{H}$ is given by 
\begin{align}
\mathcal{A} & = \mathcal{A}_M\otimes \mathcal{A}_F, \\
\mathcal{A}_M & = \mathcal{A}_+ \oplus \mathcal{A}_-.
\end{align}
Here an element $u_a$ of $\mathcal{A_+}$, which acts on $\mathcal{H}_+$ and an element $\bar{u}_a$ of 
$\mathcal{A}_-$, which acts on $\mathcal{H}_-$ are given by a chiral superfield 
expressed in the coordinate $(x^\mu_+,\theta,\bar{\theta})$ and an antichiral 
superfield expressed in the coordinate $(x^\mu_-,\theta, \bar{\theta})$, respectively, 
\begin{align}
u_a(x_+) = & \frac{1}{m_0}(\varphi_a + \sqrt{2}\theta\psi_a+\theta\theta F_a), \label{ua}\\
\bar{u}_a(x_-) = & \frac{1}{m_0}(\varphi_a^\ast+\sqrt{2}\bar{\theta}\bar{\psi}_a +\ovl{\theta\theta}F_a^\ast), 
\label{bua}
\end{align}
where we introduce a constant $m_0$ with mass dimension 1 for adjustment of the dimension.  

As for the algebra of the finite space $\mathcal{A}_F$, we assume that $\mathcal{A}_F$ is the space of $N\times N$ complex matrix functions $M_N$. 
We impose that for the particle part $Q^a$, the size $N$ of the matrix is greater than one, which will  
lead to non-abelian $U(N)$ internal symmetry and for the antiparticle part $Q_a^c$, $N$ is equal to one, 
which will lead to the abelian $U(1)$ symmetry. We note that these superfields 
$u_a(\bar{u_a}) \otimes M_N$ should not be confused with those of the functional 
space in Eqs.(\ref{Psip}) and (\ref{Psim}). As we will discuss in the next section, 
$u_a\otimes M_N$ and $\bar{u}_a\otimes M_N$ together with the Dirac operator will be the origin of the gauge supermultiplets, 
while the elements (\ref{Psip}) and (\ref{Psim}) of the functional space are the origin of the matter fields. 

The total supersymmetric Dirac operator is defined by
\begin{equation}
{\rm i}\mathcal{D}_{tot} = {\rm i}\mathcal{D}_M + \gamma_M\otimes \mathcal{D}_F. \label{DAtot}
\end{equation}
In order to specify the Dirac operator $\mathcal{D}_M$ on the Minkowskian manifold, 
we introduce the two operators $\mathcal{D}$ and $\bar{\mathcal{D}}$. 
In the coordinate $(x^\mu_-,\theta,\bar{\theta})$, the operator $\mathcal{D}$ is given by
\begin{equation}
\mathcal{D}(x_-) = -\frac{1}{4}\varepsilon^{\alpha\beta}\frac{\partial}{\partial\theta^\beta}\frac{\partial}{\partial\theta^\alpha}
=\frac{1}{4}\varepsilon^{\alpha\beta}\frac{\partial}{\partial\theta^\alpha}\frac{\partial}{\partial\theta^\beta}, \label{mD}
\end{equation}   
and in the coordinate $(x^\mu_+,\theta,\bar{\theta})$, the operator $\bar{\mathcal{D}}$ is given by 
\begin{equation}
\bar{\mathcal{D}}(x_+) = -\frac{1}{4}\varepsilon^{\dot{\alpha}{\dot{\beta}}}
\frac{\partial}{\partial\bar{\theta}^{\dot{\alpha}}}\frac{\partial}{\partial\bar{\theta}^{\dot{\beta}}}.  \label{bmD}
\end{equation} 
When we represent the chiral and antichiral superfields in the coordinate $(x_-^\mu,\theta,\bar{\theta})$ 
and $(x_+^\mu,\theta,\bar{\theta})$, respectively, 
they are given by 
\begin{align}
\Psi_+(x_-) = & \varphi_+ +\sqrt{2}\theta\psi_+ +\theta\theta F_+ +2{\rm i}\theta\sigma^\mu\bar{\theta}\partial_\mu\varphi_+
+\theta\theta\ovl{\theta\theta}\Box\varphi_+ -\sqrt{2}{\rm i}\theta\theta\partial_\mu\psi_+\sigma^\mu\bar{\theta}
\label{Psipm}, \\
\Psi_-(x_+)= & \varphi_-^\ast+\sqrt{2}\bar{\theta}\bar{\psi}_- + \overline{\theta\theta} F_-^\ast
- 2{\rm i}\theta\sigma^\mu\bar{\theta}\partial_\mu\varphi_-^\ast
+\theta\theta\ovl{\theta\theta}\Box\varphi_-^\ast+\sqrt{2}{\rm i}\ovl{\theta\theta}
\theta\sigma^\mu\partial_\mu\bar{\psi}_-,
\label{Psimp}
\end{align}
so that the left-handed and right-handed matter fields can also be expressed by  
\begin{equation}
\Phi_L(x_-)= q^a_L \otimes (\Psi_+(x_-),0)^T,\ \ \Phi_R(x_+)= q^a_R \otimes (0,\Psi_-(x_+))^T. \label{Phimp}
\end{equation}
Then the result of operation of $\mathcal{D}$ and $\bar{\mathcal{D}}$ on these fields are given by
\begin{align}
\mathcal{D}\Psi_+(x_-) = &  
F_+ + \sqrt{2}\bar{\theta}{\rm i}\bar{\sigma}^\mu\partial_\mu\psi_+ +\overline{\theta\theta}\Box\varphi_+ ,\ \ 
\mathcal{D}\Phi_L = q^a_L\otimes(0,\mathcal{D}\Psi_+(x_-))^T,
\\
\bar{\mathcal{D}}\Psi_-(x_+) = &  
F_-^\ast + \sqrt{2}\theta{\rm i}\sigma^\mu\partial_\mu\bar{\psi}_- +\theta\theta\Box\varphi_-^\ast,\ \ 
\bar{\mathcal{D}}\Phi_R= q_R^a\otimes(\bar{\mathcal{D}}\Psi_-(x_+),0)^T,
\end{align}
where we note that 
\begin{equation}
\mathcal{D}\mathcal{H}_+ \subset \mathcal{H}_-,\ \ \bar{\mathcal{D}}\mathcal{H}_- \subset \mathcal{H}_+.
\end{equation}
The Dirac operator on the manifold is defined on the basis (\ref{HM}) by
\begin{equation}
{\rm i}\mathcal{D}_M = 
\begin{pmatrix}
0 & \bar{\mathcal{D}} \\
\mathcal{D} & 0 
\end{pmatrix}. \label{DM}
\end{equation} 

The Dirac operator on the finite space is defined on the basis (\ref{Qa}) by
\begin{equation}
\mathcal{D}_F =
\begin{pmatrix}
m & 0 \\
0 & m^\dagger
\end{pmatrix},  \label{DF}
\end{equation}
where $m$ and $m^\dagger $ are mass matrix with respect to the family index.

The supersymmetric invariant product in $\mathcal{H}_M$ is defined as follows: \\
In $\mathcal{H}_+$,
\begin{equation}
(\Psi_-,\Psi_+^\prime)_s = \int_M d^4 x d^2\theta\delta(\bar{\theta}) \Psi_-^\dagger \Psi_+^\prime, \label{suprop}
\end{equation}
and in $\mathcal{H}_-$, 
\begin{equation}
(\Psi_+,\Psi_-^\prime)_s = \int_M d^4 x d^2\ovl{\theta}\delta(\theta)\Psi_+^\dagger \Psi_-^\prime. \label{suprom}
\end{equation}
For example, when we couple Eqs.(\ref{suprop}),(\ref{suprom}) with elements $Q^a$ of $\mathcal{H}_F$, we 
obtain that 
\begin{align}
(\Phi_L,\mathcal{D}\Phi_L)_s = & \int_M d^4 x d^2\bar{\theta}\delta(\theta) q^a_L\otimes(0,\Psi_+^\dagger) \mathcal{D} \Phi_L
= \int d^4 x (\varphi_L^\ast \Box\varphi_L -{\rm i}\bar{\psi}_L\bar{\sigma}^\mu\partial_\mu \psi+F_L^\ast F_L),
\label{PLDPL} \\
(\Phi_R,\bar{\mathcal{D}}\Phi_R)_s = & \int_M d^4 x d^2\theta \delta(\bar{\theta})q^a_R\otimes(\Psi_-^\dagger,0) \bar{\mathcal{D}} \Phi_R
= \int d^4 x (\varphi_R^\ast \Box\varphi_R -{\rm i}\bar{\psi}_R\sigma^\mu\partial_\mu \psi_R+F_R^\ast F_R).
\label{PRDPR}
\end{align}
The expressions (\ref{PLDPL}) and (\ref{PRDPR}) give the kinetic terms of matter fields without gauge fields.
\section{Internal fluctuation and vector supermultiplet}
In the supersymmetric counterpart of the NCG, the vector supermultiplet is to be introduced as the internal 
fluctuation of the Dirac operator in (\ref{DAtot}), which is given by
\begin{equation}
{\rm i}\mathcal{D}_{tot} \rightarrow {\rm i}\tilde{\mathcal{D}}_{tot}={\rm i}\mathcal{D}_{tot} 
+ V + JVJ^{-1},\ V = \sum_a U_a^\prime [{\rm i}\mathcal{D}_{tot},U_a],\ U_a\in A, \label{ttD}
\end{equation} 
where $J=J_M\otimes J_F$. Since $U_a$ is a complex constant for the space of $Q^a_c$, its contribution to the fluctuation vanishes. 
But the third term $JVJ^{-1}$ of the r.h.s. carries the same fluctuation by $N\times N$ complex matrix functions $M_N$ in the 
space of $Q^a $ to the space of $Q_c^a$.

The algebra $\mathcal{A}_M$ is 
a sum of $\mathcal{A}_+$ and $\mathcal{A}_-$, so we prepare two sets of elements, $\Pi_+$ and $\Pi_-$:
\begin{align}
\Pi_+ & = \{u_a:a=1,2,\cdots n\} \subset \mathcal{A}_+\otimes \mathcal{A}_F, \\
\Pi_- & = \{\bar{u}_a:a=1,2,\cdots n\} \subset\mathcal{A}_-\otimes \mathcal{A}_F,
\end{align} 
where $u_a$ and $\bar{u}_a$ are given in (\ref{ua}) and (\ref{bua}). 
Since the product of chiral(antichiral) superfields is again a chiral(antichiral) superfield, the 
elements of $\Pi_+$($\Pi_-$) are chosen such that products of two or more $u_as(\bar{u}_as) $ 
do not belong to $\Pi_+$($\Pi_-$) any more.

We shall define the following components of vector superfields as the bilinear form of the 
two component functions in $u_a\in \Pi_+$ and $\bar{u}_a\in \Pi_- $:
\begin{align}
m_0^2 C & = \sum_a c_a\varphi_a^\ast \varphi_a, \label{m02C}\\
m_0^2 \chi_\alpha & = -{\rm i}\sqrt{2}\sum_ac_a\varphi_a^\ast\psi_{a\alpha},\\
m_0^2(M+{\rm i}N) & =-2{\rm i}\sum_a c_a\varphi^\ast_aF_a,\\
m_0^2A_\mu & = -{\rm i}\sum_a c_a[(\varphi_a^\ast\partial_\mu\varphi_a-\partial_\mu\varphi_a^\ast\varphi_a)
-{\rm i}\bar{\psi}_{a\dot{\alpha}}\bar{\sigma}^{\dot{\alpha}\alpha}_\mu\psi_{a\alpha}], \\
m_0^2\lambda_\alpha & = \sqrt{2}{\rm i}\sum_a c_a(F_a^\ast\psi_{a\alpha}
-{\rm i}\sigma^\mu_{\alpha\dot{\alpha}}\bar{\psi}_a^{\dot{\alpha}}\partial_\mu\varphi_a), \\
m_0^2 D & = \sum_a c_a[2F_a^\ast F_a -2(\partial^\mu\varphi_a^\ast\partial_\mu\varphi_a) \nonumber \\
& +{\rm i}\{\partial_\mu\bar{\psi}_{a\dot{\alpha}}\bar{\sigma}^{\mu\dot{\alpha}\alpha}\psi_{a\alpha}
-\bar{\psi}_{a\dot{\alpha}}\bar{\sigma}^{\mu\dot{\alpha}\alpha}\partial_\mu\psi_{a\alpha}\}], \label{m02D}
\end{align}
where $c_a$ are the real coefficients. These component fields have the transformation property of the vector 
supermultiplets.

When we define the vector superfield(\ref{m02C})$\sim$ (\ref{m02D}), there is an ambiguity 
due to the choice of the algebraic elements and we can redefine $C,\xi_a,M$ and $N$ to be zero. 
The choice of the vector supermultiplet is the Wess--Zumino gauge. 
This gauge is realized in (\ref{m02C})$\sim$ (\ref{m02D}) by the 
following condition:
\begin{align}
\sum_a c_a\varphi_a^\ast \varphi_a & = 0, \nonumber \\
\sum_a c_a\varphi_a^\ast \psi_a^\alpha & =0, \label{WessZumino}\\
\sum_a c_a\varphi_a^\ast F_a & =0.\nonumber 
\end{align}

The vector supermultiplets 
$A_\mu$,$D$, $\lambda_\alpha$ are also $N\times N$ complex matrix functions and 
parametrized by
\begin{align}
A_\mu(x) & = \sum_{l=0}^{N^2-1} A_\mu^l(x)\frac{T_l}{2}, \label{Amu} \\
D(x) & = \sum_{l=0}^{N^2-1} D^l(x)\frac{T_l}{2}, \label{Dx}\\       
\lambda_\alpha(x) & = \sum_{l=0}^{N^2-1} \lambda_\alpha^l(x)\frac{T_l}{2}. \label{lambda}
\end{align}
Here, $T^l$ are basis of generators which belong to fundamental representation of the Lie algebra associated 
with Lie group $U(N)$ and normalized as follows:
\begin{equation}
{\rm Tr}(T_aT_b) =2\delta_{ab}.
\end{equation}  
Since $A^\mu(x)$ and $D(x)$ are hermitian so that $A_\mu^l(x)$ and $D^l(x)$ are real functions. On the other hand, 
$\lambda_\alpha^l(x)$ are complex functions.

We denote the supersymmetric Dirac operator modified by the fluctuation as follows:  
\begin{equation}
\tilde{\mathcal{D}}_M = -{\rm i}\begin{pmatrix}
0 & \tilde{\bar{\mathcal{D}}}\\
\tilde{\mathcal{D}} & 0
\end{pmatrix}.
\label{tDM}
\end{equation}
We consider the fluctuation due to $u_a\in \Pi_+$ and $\bar{u}_a\in \Pi_-$. 
In the base of (\ref{HM}), we take $U_a,U_a^\prime$ in (\ref{ttD}) as follows:
\begin{equation}
U_a =  \sqrt{-2c_a}\begin{pmatrix}
u_a & 0\\
0 & 0
\end{pmatrix},\ \ 
U_a^\prime =\sqrt{-2c_a}
\begin{pmatrix}
0 & 0\\
0 & \bar{u}_a \label{UaUa}
\end{pmatrix}.
\end{equation}
Then, the contribution of $U_a$ and $U_a^\prime$ to $\tilde{\mathcal{D}}$ is given by 
the following form:
\begin{equation}
V_{\mathcal{D}}  = -2 \sum_a c_a \bar{u}_a
[{\rm i}\mathcal{D}_M,u_a] 
= -2\sum_ac_a \bar{u}_a\mathcal{D}u_a, \label{V1}
\end{equation}
and when we take $U_a$, $U_a^\prime$ as follows:
\begin{equation}
U_a =\sqrt{2c_a}\begin{pmatrix}
0 & 0\\
0 & \bar{u}_a 
\end{pmatrix},\ \ 
U_a^\prime = \sqrt{2c_a}\begin{pmatrix}
u_a & 0\\
0 & 0
\end{pmatrix},  \label{bUaUa}
\end{equation}
the contribution to $\tilde{\bar{D}} $ is given by
\begin{equation}
V_{\bar{\mathcal{D}}}  = 2\sum_a c_a u_a[{\rm i}\mathcal{D}_M,\bar{u}_a] 
= 2\sum_ac_a u_a\bar{\mathcal{D}}\bar{u}_a. \label{bV1}
\end{equation}
We shall calculate in the Wess--Zumino gauge. Using the definition of the 
vector supermultiplet given by (\ref{m02C})$\sim$ (\ref{m02D}), 
we obtain the following result:
\begin{align}
-\frac{V_{\mathcal{D}}}{2}= & \sum_{a}c_a\bar{u}_a\mathcal{D}u_a = \overline{\theta\theta}\frac{1}{2}(D+{\rm i}(\partial^\mu A_\mu))+{\rm i}\overline{\lambda\theta}
\nonumber \\
& -\frac{1}{2}(D+{\rm i}(\partial^\mu A_\mu))
\overline{\theta\theta}\theta^\alpha\frac{\partial}{\partial\theta^\alpha}
+\frac{{\rm i}}{2}\lambda^\alpha\overline{\theta\theta}\frac{\partial}{\partial\theta^\alpha}
-{\rm i}\overline{\lambda\theta}\theta^\alpha\frac{\partial}{\partial\theta^\alpha}
-\frac{1}{2}A^\mu\bar{\theta}_{\dot{\alpha}}\sigma_\mu^{\dot{\alpha}\alpha}\frac{\partial}{\partial\theta^\alpha}
 \nonumber \\
& +\left(\frac{1}{2}(D+{\rm i}(\partial^\mu A_\mu))\theta\theta\overline{\theta\theta}
-A_\mu \theta\sigma^\mu\bar{\theta}
+{\rm i}\bar{\lambda}\bar{\theta}\theta\theta
-{\rm i}\overline{\theta\theta}\theta\lambda
\right)\mathcal{D},
\label{buDu}
\end{align}
in the coordinate $(x_-^\mu,\theta,\bar{\theta})$ and 
\begin{align}
\frac{V_{\bar{\mathcal{D}}}}{2}= & \sum_a c_a u_a\bar{\mathcal{D}}\bar{u}_a =  
\theta\theta\frac{1}{2}(\bar{\mathcal{D}}-{\rm i}(\partial^\mu A_\mu))-{\rm i}\theta\lambda \nonumber \\
& -\frac{1}{2}(\bar{\mathcal{D}}-{\rm i}(\partial^\mu A_\mu))
\theta\theta\bar{\theta}^{\dot{\alpha}}\frac{\partial}{\partial \bar{\theta}^{\dot{\alpha}}}
+\frac{\rm i}{2}\bar{\lambda}_{\dot{\alpha}}\theta\theta\varepsilon^{\dot{\alpha}\dot{\beta}}
\frac{\partial}{\partial \bar{\theta}^{\dot{\beta}}}
+{\rm i}\lambda\theta\bar{\theta}^{\dot{\alpha}}\frac{\partial}{\partial \bar{\theta}^{\dot{\alpha}}}
-\frac{1}{2}A^\mu\bar{\sigma}_\mu^{\dot{\alpha}\alpha}\theta_\alpha\frac{\partial}{\partial \bar{\theta}^{\dot{\alpha}}} \nonumber \\
& +\left(\frac{1}{2}(D-i(\partial_\mu A^\mu))\theta\theta\overline{\theta\theta}
-{\rm i}\theta\lambda\overline{\theta\theta}
+{\rm i}\overline{\lambda\theta}\theta\theta
-\theta\sigma^\mu\bar{\theta} A_\mu \right)\bar{\mathcal{D}},
\label{ubDbu}
\end{align}
in the coordinate $(x_+^\mu,\theta,\bar{\theta})$.

Let us consider the fluctuations due to the product of two elements in $\Pi_+$ and $\Pi_-$ 
expressed by $u_{ab}=u_au_b$ and $\bar{u}_{ab}=\bar{u}_a\bar{u}_b$. We replace 
$U_a$, $U_a^\prime$ in Eq.(\ref{UaUa}), (\ref{bUaUa}) 
with $U_a U_b$, $U_a^\prime U_b^\prime$ and 
obtain them as follows:

\begin{equation}
\sum_{a,b}c_ac_b\bar{u}_{ab}\mathcal{D}u_{ab}
= -\frac{1}{2}A^\mu A_\mu\overline{\theta\theta}(1-\theta^\alpha\frac{\partial}{\partial\theta^\alpha}
+\theta\theta)\mathcal{D},
\label{buuDuu}
\end{equation}
in the coordinate $(x_-,\theta,\bar{\theta})$ and 
\begin{equation}
\sum_{a,b}c_ac_b u_{ab}\bar{\mathcal{D}}\bar{u}_{ab}= 
-\frac{1}{2}A^\mu A_\mu\theta\theta(1-\bar{\theta}^{\dot{\alpha}}
\frac{\partial}{\partial \bar{\theta}^{\dot{\alpha}}}+\overline{\theta\theta})\bar{\mathcal{D}},
\label{uabbDbuab}
\end{equation}
in the coordinate $(x_+,\theta,\bar{\theta})$. The fluctuation due to higher order products of 
$u_a$ or $\bar{u}_a$ such as $u_{abc}=u_au_bu_c$ or $\bar{u}_{abc}= \bar{u}_a\bar{u}_b\bar{u}_c$ 
vanishes due to the Wess--Zumino gauge condition.

The Dirac operator with fluctuation denoted by (\ref{tDM})
is finally obtained. Using (\ref{buDu}),(\ref{buuDuu}), the fluctuated $\mathcal{D}$ is given 
in the coordinate $(x_-,\theta,\bar{\theta})$ by 
\begin{align}
\lefteqn{\tilde{\mathcal{D}}(x_-)=\mathcal{D}-2\bar{u}_a\mathcal{D}u_a
+2\bar{u}_{ab} \mathcal{D}u_{ab}}\nonumber \\
= & 
\mathcal{D}-(D+{\rm i}(\partial_\mu A^\mu)+A^\mu A_\mu)\overline{\theta\theta}-2{\rm i}\overline{\lambda\theta}
+ (D+{\rm i}(\partial_\mu A^\mu)+A^\mu A_\mu)\overline{\theta\theta}\theta^\alpha\frac{\partial}{\partial \theta^\alpha} 
\nonumber \\
& -{\rm i}\lambda^\alpha\overline{\theta\theta}\frac{\partial}{\partial \theta^\alpha}
+2{\rm i}\overline{\lambda\theta}\theta^\alpha\frac{\partial}{\partial \theta^\alpha}
+A^\mu\bar{\theta}_{\dot{\alpha}}\bar{\sigma}_\mu^{\dot{\alpha}\alpha}\frac{\partial}{\partial \theta^\alpha}
\nonumber \\
& -(D+{\rm i}(\partial_\mu A^\mu)+A^\mu A_\mu)\theta\theta\overline{\theta\theta}\mathcal{D}
+2A_\mu\theta\sigma^\mu\bar{\theta}\mathcal{D}-2{\rm i}\overline{\lambda\theta}\theta\theta\mathcal{D}
+2{\rm i}\lambda\theta\overline{\theta\theta}\mathcal{D},
\label{tildeD}
\end{align}
and from (\ref{ubDbu}),(\ref{uabbDbuab}), the fluctuated $\bar{\mathcal{D}}$ is given 
in the coordinate$(x_+,\theta,\bar{\theta})$ by
\begin{align}
\lefteqn{\tilde{\bar{\mathcal{D}}}(x_+) = 
\bar{\mathcal{D}}+2u_a\bar{\mathcal{D}}\bar{u}_a +2u_{ab}\bar{\mathcal{D}}\bar{u}_{ab}} \nonumber \\
= & \bar{\mathcal{D}}+(D-{\rm i}(\partial^\mu A_\mu) - A^\mu A_\mu)\theta\theta -2{\rm i}\theta\lambda
-(D-{\rm i}(\partial^\mu A_\mu) - A^\mu A_\mu)\theta\theta\bar{\theta}^{\dot{\alpha}}
\frac{\partial}{\partial \bar{\theta}^{\dot{\alpha}}} \nonumber\\
& +{\rm i}\bar{\lambda}_{\dot{\alpha}}\theta\theta \varepsilon^{\dot{\alpha}\dot{\beta}}
\frac{\partial}{\partial \bar{\theta}^{\dot{\beta}}}+2{\rm i}\lambda\theta\bar{\theta}^{\dot{\alpha}}
\frac{\partial}{\partial \bar{\theta}^{\dot{\alpha}}} 
-A_\mu\bar{\sigma}^{\dot{\alpha}\alpha}\theta_\alpha
\frac{\partial}{\partial \bar{\theta}^{\dot{\alpha}}} \nonumber \\
& +(D-{\rm i}(\partial^\mu A_\mu) - A^\mu A_\mu)\theta\theta\overline{\theta\theta}\bar{\mathcal{D}}
-2A_\mu\theta\sigma^\mu\bar{\theta}\bar{\mathcal{D}}
-2{\rm i}\theta\lambda\overline{\theta\theta}\bar{\mathcal{D}}
+2{\rm i}\overline{\lambda\theta}\theta\theta\bar{\mathcal{D}}.
\label{tildebD}
\end{align}
As for the Dirac operator on the finite space, we assume that $\mathcal{D}_F$ in Eq.(\ref{DF}) has no internal degrees of 
freedom, so the fluctuation for it does not arise.
\section{Spectral action principle and super Yang-Mills action}
The action of NCG models is obtained by the spectral action principle expressed given in 
Eq.(\ref{totalaction0}).

Let us see the counterpart of the first term in Eq.(\ref{totalaction0}) which is in our supersymmetric case 
the part of the spectral action for the matter particles and their superpartners. The modified total Dirac operator 
on the basis $\Phi_L \oplus \Phi_R$ is given with the expressions (\ref{DF}),(\ref{tDM}) by
\begin{equation}
{\rm i}\tilde{\mathcal{D}}_{tot} = {\rm i}\tilde{\mathcal{D}}_M\otimes \bf{1}_F + \gamma_M \otimes \mathcal{D}_F. 
\label{Dtot}
\end{equation} 
The action for the matter fields is expressed by the bilinear form of supersymmetric invariant product 
(\ref{suprop}) and (\ref{suprom}) with the total Dirac operator as follows: 
\begin{align}
I_{matter} & = \left(\Phi_L+\Phi_R,{\rm i}\mathcal{D}_{tot}(\Phi_L+\Phi_R)\right)_s \nonumber \\
& = \left(\Phi_L+\Phi_R,{\rm i}\mathcal{D}_M(\Phi_L+\Phi_R)\right)_s
+\left(\Phi_L+\Phi_R,\gamma_M\otimes \mathcal{D}_F (\Phi_L+\Phi_R)\right)_s \nonumber \\
& = (\Phi_L,\tilde{\mathcal{D}}\Phi_L)_s+(\Phi_R,\tilde{\ovl{\mathcal{D}}}\Phi_R)_s
+(\Phi_L,{\rm i}m^\dagger\Phi_R)_s -(\Phi_R,{\rm i}m\Phi_L)_s. \label{Imatter}
\end{align}
Using the left-handed and right-handed superfields in (\ref{Phimp})  with 
fluctuated Dirac operators in (\ref{tildeD}) and (\ref{tildebD}), 
the kinetic parts of the matter particles are obtained by 
\begin{align}
I_L = & (\Phi_L,\tilde{\mathcal{D}}\Phi_L)_s 
\nonumber \\
= & \int_M d^4x\left(
\varphi_L^\ast(D^\mu D_\mu-D)\varphi_L-{\rm i}\bar{\psi}_L\bar{\sigma}^\mu D_\mu\psi_L
+F_L^\ast F_L -\sqrt{2}{\rm i}(\varphi_l^\ast\lambda\psi_L-\bar{\psi}\bar{\lambda}\psi_L)
\right),
\label{PLtDPL}
\end{align}
and 
\begin{align}
I_R = & (\Phi_R,\tilde{\bar{\mathcal{D}}}\Phi_R)_s
\nonumber \\
= & \int_M d^4x\left(
\varphi_R^\ast(D^\mu D_\mu+D)\varphi_R -{\rm i}\bar{\psi}_R \sigma^\mu D_\mu\psi_R + F_R^\ast F_R
-\sqrt{2}{\rm i}(\varphi_R^\ast\bar{\lambda}\psi_R-\bar{\psi}_R\lambda\varphi_R)
\right). \label{PRtDPR}
\end{align}
As for the mass terms, we redefine the phase of $\Phi_L$ as $\Phi_L \rightarrow i\Phi_L$ in the last two terms of (\ref{Imatter}) 
and we have
\begin{align}
I_{\rm mass} & = (\Phi_R,m\Phi_L) + {\rm h.c.} \nonumber \\
& =\int_M {\rm d}^4x
[\varphi_R^\ast mF_L +F_R^\ast m\varphi_L-\bar{\psi}_R^\alpha m \psi_{L\alpha} +{\rm h.c.}]. \label{Imass}
\end{align}  

Now, let us see the counterpart of the second term in Eq.(\ref{totalaction0}) 
and derive the action of the super Yang-Mills theory. In our noncommutative geometric approach to SUSY model, 
the action for the vector supermultiplet will be obtained by the coefficients of heat 
kernel expansion of elliptic operator $P$:
\begin{equation}
Tr_{L^2}f(P) \simeq \sum_{n\geq 0}c_na_n(P),\label{hke}
\end{equation}
where $f(x)$ is an auxiliary smooth function on a smooth compact Riemannian manifold without boundary of 
dimension four similar to the non-supersymmetric case.
Since the contribution to $P$ from the antiparticles is 
the same as that of the particles, we consider only the contribution from the particles. 
Then the elliptic operator $P$ in our case is given by the square of the Dirac operator ${\rm i}\tilde{\mathcal{D}}_{tot}$ 
in (\ref{Dtot}). 
We expand the operator $P$ into the following form:
\begin{equation}
P = (i\mathcal{D}_{tot})^2 
= \eta^{\mu\nu}\partial_\mu\partial_\nu +\mathbb{A}^\mu\partial_\mu +\mathbb{B}. \label{P}
\end{equation}
The heat kernel coefficients $a_n$ in Eq.
(\ref{hke}) are found in \cite{gilkey}. 
They vanish for odd $n$, and the first three $a_n$'s for 
even $n$ in our model are given by
\begin{align}
a_0(P) & =\frac{1}{16\pi^2}\int_M
{\rm d}^4x~
 {\rm Str}(\mathbb{I}), \label{a0} \\
a_2(P) & =\frac{1}{16\pi^2}\int_M
{\rm d}^4x~
{\rm Str}(\mathbb{E}), \label{a2} \\
a_4(P) & =\frac{1}{32\pi^2}\int_M
{\rm d}^4x~
 {\rm Str}(\mathbb{E}^2 \label{a4}
+\frac{1}{3}\mathbb{E}_{;\mu}^\mu+\frac{1}{6}\Omega^{\mu\nu}\Omega_{\mu\nu}), 
\end{align}
where $\mathbb{E}$ and the bundle curvature $\Omega^{\mu\nu}$ are defined as follows:
\begin{align}
\mathbb{E} & = \mathbb{B} -(\partial_\mu \omega^\mu+\omega_\mu\omega^\mu), \label{E}\\
\Omega^{\mu\nu} & = \partial^\mu\omega^\nu -\partial^\nu\omega^\mu+[\omega^\mu,\omega^\nu], \label{Omega}\\
\omega^\mu & =\frac{1}{2}\mathbb{A}^\mu. \label{omega}
\end{align}
In Eqs.(\ref{a0}-\ref{a4}), Str denotes the trace over the indices of internal 
degrees of freedom and supertrace over the spin degrees of freedom. 

The coefficients $c_n$ in (\ref{hke}) depend on the functional form of $f(x)$. If $f(x)$ is flat near 0, 
it turns out that $c_{2k}=0$ for $k\geq 3$
 and the heat kernel expansion terminates at 
$n=4$ \cite{connes8}.  

In order to represent $\mathbb{E}$, $\omega^\mu$, $\mathbb{E}^2$, $\Omega_{\mu\nu}\Omega^{\mu\nu}$ 
in the superspace coordinate, we introduce some operators on the functional space $\mathcal{H}_\pm$. 
Hereafter, we represent operators acting on $\mathcal{H}_+$ in the coordinate $(x_+^\mu,\theta,\bar{\theta})$ 
and operators acting on $\mathcal{H}_-$ in the coordinate $(x_-^\mu,\theta,\bar{\theta})$.  
Let $f_+(x_+)$ and $f_-(x_-)$ be an element in 
$\mathcal{H}_+$ and an element in $\mathcal{H}_-$, respectively, 
\begin{equation}
f_+ (x_+) = f_0 +\sqrt{2}\theta^\alpha f_{1\alpha}+\theta\theta f_2,\ \ 
f_- (x_-) = f_0^\ast +\sqrt{2}\bar{\theta}_{\dot{\alpha}} \bar{f}_1^{\dot{\alpha}}
 +\ovl{\theta\theta} f_2^\ast.
\end{equation} 
We define following operators: 
\begin{equation}
\mathcal{D}_+ = 
\frac{1}{4}\varepsilon^{\alpha\beta}\frac{\partial}{\partial\theta^\alpha}\frac{\partial}{\partial\theta^\beta},\ 
\hat{f}_0 =  \mathcal{D}_+\theta\theta,\ \hat{f}_{1\alpha}=-\sqrt{2}\mathcal{D}_+\theta_\alpha,\ 
\hat{f}_2=\mathcal{D}_+,
\label{fp02} 
\end{equation}
which act on $f_+$ and 
\begin{equation}
\mathcal{D}_- = -\frac{1}{4}\varepsilon^{\dot{\alpha}{\dot{\beta}}}
\frac{\partial}{\partial\bar{\theta}^{\dot{\alpha}}}\frac{\partial}{\partial\bar{\theta}^{\dot{\beta}}},\ 
\hat{f}_0^\ast =  \mathcal{D}_- \ovl{\theta\theta},\  
\hat{\bar{f}}_1^{\dot{\alpha}}=-\sqrt{2}\mathcal{D}_- \bar{\theta}^{\dot{\alpha}},\ 
\hat{f}_2^\ast = \mathcal{D}_- ,
\label{fm02}
\end{equation}
which act on $f_-$. 
These operators extract the components of the superfileds $f_0,f_{1\alpha},f_2,$ $f_0^\ast, \bar{f}_1^{\dot{\alpha}}, f_2^\ast$ from 
$f_+$ and $f_-$ and satisfy following equations: 
\begin{equation}
\hat{f}_0 f_+ =f_0,\ \hat{f}_{1\alpha}f_+ = f_{1\alpha}, \hat{f}_2f_+ =f_2,\ 
\hat{f}_0^\ast f_- =f_0^\ast,\ \hat{\bar{f}}_1^{\dot{\alpha}}f_- =  \bar{f}_1^{\dot{\alpha}},\ 
\hat{f}_2^\ast f_- = f_2^\ast. 
\end{equation}
We refer to operators in (\ref{fp02}), (\ref{fm02}) as "extracting operators". 
With the extracting operators, we can also make identity operators on $\mathcal{H}_+$ and $\mathcal{H}_-$ as follows:
\begin{equation}
\mathbb{I}_+=\hat{f}_0+\sqrt{2}\theta^\alpha\hat{f}_{1\alpha}+\theta\theta\hat{f}_2,\ 
\mathbb{I}_-=\hat{f}_0^\ast+\sqrt{2}\bar{\theta}_{\dot{\alpha}}\hat{\bar{f}}_1^{\dot{\alpha}}
+\ovl{\theta\theta}\hat{f}_2^\ast \label{I},
\end{equation}
which satisfy $\mathbb{I}_+f_+ = f_+$, $\mathbb{I}_-f_- = f_-$.

We define the supertrace Str necessary to obtain the heat kernel expansion coefficients as follows:
for an operator on $\mathcal{H}_+$, it is given by
\begin{equation}
{\rm Str}_+ = \hat{f}_0\frac{\delta}{\delta  \hat{f}_0 }
+ \hat{f}_{1\alpha}\frac{\delta}{\delta  \hat{f}_{1\alpha} }
+  \hat{f}_2\frac{\delta}{\delta  \hat{f}_2 }, \label{Strp}
\end{equation}
and for an operator on $\mathcal{H}_-$, it is  given by 
\begin{equation}
{\rm Str}_- = \hat{f}_0^\ast\frac{\delta}{\delta  \hat{f}_0^\ast }
+ \hat{\bar{f}}_{1\alpha}\frac{\delta}{\delta  \hat{\bar{f}}_{1\alpha} }
+  \hat{f}_2^\ast\frac{\delta}{\delta  \hat{f}_2^\ast }. \label{Strm}
\end{equation}

Let $\hat{\mathcal{A}}$ be an operator on $\mathcal{H}_+$ and be expressed by 
\begin{align}
\hat{\mathcal{A}} = &
\mathcal{A}_0\hat{f}_0 + (\cdots)^\alpha\hat{f}_{1\alpha} + (\cdots)\hat{f}_2 \nonumber \\
& +\sqrt{2}\theta^\alpha(\cdots)_\alpha\hat{f}_0+\sqrt{2}\theta^\beta\mathcal{A}_{1\beta}^{\ \ \alpha}\hat{f}_{1\alpha}
+\sqrt{2}\theta^\alpha(\cdots)_\alpha \hat{f}_2 \nonumber\\
& +\theta\theta(\cdots)\hat{f}_0+\theta\theta(\cdots)^\alpha\hat{f}_{1\alpha}+\theta\theta\mathcal{A}_2\hat{f}_2.
\end{align}
The supertrace $\rm{Str}_+$ of $\hat{\mathcal{A}}$ is given by
\begin{align}
{\rm Str}_+(\hat{\mathcal{A}}) = & \hat{f}_0(\mathcal{A}_0 + \sqrt{2}\theta^\alpha(\cdots)_\alpha+\theta\theta(\cdots)) \nonumber\\
& + \hat{f}_{1\alpha}(-(\cdots)^\alpha -\sqrt{2}\theta^\beta\mathcal{A}_{1\beta}^{\ \ \alpha} -\theta\theta(\cdots)^\alpha)\nonumber \\
& +\hat{f}_2( (\cdots)+\sqrt{2}\theta^\alpha(\cdots)_\alpha+\theta\theta\mathcal{A}_2) \nonumber\\
= & \mathcal{A}_0 -\mathcal{A}_{1\alpha}^{\ \ \alpha}+\mathcal{A}_2, 
\end{align}
while for an operator on $\mathcal{H}_-$ expressed by
\begin{align}
\hat{\bar{\mathcal{A}}} = &
\mathcal{A}_0^\ast\hat{f}_0^\ast + (\cdots)^{\dot{\alpha}}\hat{\bar{f}}_{1\dot{\alpha}} + (\cdots)\hat{f}_2^\ast \nonumber \\
& +\sqrt{2}\bar{\theta}_{\dot{\alpha}}(\cdots)^{\dot{\alpha}}\hat{f}_0^\ast+
\sqrt{2}\bar{\theta}_{\dot{\alpha}}\mathcal{A}_{1\ \dot{\beta}}^{\ \dot{\alpha}}\hat{\bar{f}}_1^{\dot{\beta}}
+\sqrt{2}\bar{\theta}_{\dot{\alpha}}(\cdots)^{\dot{\alpha}} \hat{f}_2 \nonumber\\
& +\ovl{\theta\theta}(\cdots)\hat{f}_0^\ast+\ovl{\theta\theta}(\cdots)_{\dot{\alpha}}\hat{\bar{f}}_1^{\dot{\alpha}}
+\theta\theta\mathcal{A}_2^\ast \hat{f}_2^\ast,
\end{align}
the supertrace ${\rm Str}_-$ of $\hat{\bar{\mathcal{A}}}$ is given by
\begin{equation}
{\rm Str}_- (\hat{\bar{\mathcal{A}}}) = \mathcal{A}_0^\ast-\mathcal{A}_{1\ \dot{\alpha}}^{\ \dot{\alpha}}+\mathcal{A}_2^\ast.
\end{equation}
Particularly, ${\rm Str}_+(\mathbb{I}_+)={\rm Str}_-(\mathbb{I}_-)=1-2+1=0$ is established. 

Let us start the investigation of the elliptic operator $P=(i\tilde{D}_{tot})^2$.  
In the contribution of $P$ to the spectral action, the terms including 
$\mathcal{D}_F$ vanish since $\mathcal{D}_M$ anticommutes with $\gamma_M$ and 
\begin{equation}
{\rm Str}(\gamma_M^2) =-{\rm Str}(\mathbb{I}_+)-{\rm Str}(\mathbb{I}_-) =0.
\end{equation} 
Thus, we may consider $(i\mathcal{D}_M)^2$ as the operator $P$, 
\begin{equation}
P = 
\begin{pmatrix}
P_+ & 0\\
0 & P_-
\end{pmatrix}
=({\rm i}\tilde{\mathcal{D}}_M)^2 
=\begin{pmatrix}
\tilde{\bar{\mathcal{D}}} \tilde{\mathcal{D}} & 0\\
0 &  \tilde{\mathcal{D}}\tilde{\bar{\mathcal{D}}}
\end{pmatrix}.
\end{equation}

Here, we represent $P_+=\tilde{\bar{\mathcal{D}}} \tilde{\mathcal{D}}$ and 
$P_- = \tilde{\mathcal{D}}\tilde{\bar{\mathcal{D}}} $ in terms of the extracting operators by the following way: \\
At first, we operate $P_+$ on $\Psi_+ \in \mathcal{H}_+$ and $P_-$ on $\Psi_- \in \mathcal{H}_-$. 
In the process of these operations, we have to adequately switch the superspace coordinate 
which represents superfields and operators which appear there. 
As we calculate, for example, the part $\tilde{\mathcal{D}}\Psi_+$ in $P_+\Psi_+$, 
since $\tilde{\mathcal{D}}$ of (\ref{tildeD}) is represented by 
$(x_-,\theta,\bar{\theta})$, $\Psi_+$ should also be represented by $(x_-,\theta,\bar{\theta})$ 
like as (\ref{Psipm}). As we operate successively $\tilde{\bar{\mathcal{D}}}$ on the above result 
$\Psi_-^\prime =\tilde{\mathcal{D}}\Psi_+ \in \mathcal{H}_-$, since 
$\tilde{\bar{\mathcal{D}}}$ of (\ref{tildebD}) is represented by $(x_+,\theta,\bar{\theta})$, 
we should execute the operation after re-expressing  $\Psi_-^\prime$ using the coordinate 
$(x_+,\theta,\bar{\theta})$ like as (\ref{Psimp}).\\   
Secondly, in the result of the above operations, we replace the components of $\Psi_+$, $(\varphi_+,\psi_{+\alpha},F_+)$ with the extracting operators 
$(\hat{f}_0, \hat{f}_{1\alpha},\hat{f}_2)$ and the components of 
$\Psi_-$, $(\varphi_-^\ast,\bar{\psi}_-^{\dot{\alpha}},F_-^\ast)$ with 
$(\hat{f}_0^\ast, \hat{\bar{f}}_1^{\dot{\alpha}},\hat{f}_2^\ast) $.
Then, we obtain the representation of  $P_+$ on $\mathcal{H}_+$ as follows:
\begin{align}
P_+ = &  
\mathcal{D}^\mu\mathcal{D}_\mu\mathbb{I}_+-D\hat{f}_0-{\rm i}\sqrt{2}\lambda^\alpha \hat{f}_{1\alpha} \nonumber \\
& +\sqrt{2}\theta^\alpha\left((\sqrt{2}\sigma^\mu_{\alpha\dot{\alpha}}
(\mathcal{D}_\mu\bar{\lambda}^{\dot{\alpha}})+\bar{\lambda}^{\dot{\alpha}}\mathcal{D}^\mu)\hat{f}_0
+{\rm i}\sigma^{\mu\nu\ \beta}_{\ \ \alpha}F_{\mu\nu}\hat{f}_{1\beta}
-{\rm i}\sqrt{2}\lambda_\alpha f_2
\right) \nonumber \\
& +\theta\theta\left(
  -2\bar{\lambda}_{\dot{\alpha}}\bar{\lambda}^{\dot{\alpha}}\hat{f}_0 
+\sqrt{2}\bar{\lambda}_{\dot{\alpha}}\bar{\sigma}^{\mu\dot{\alpha}\beta}\mathcal{D}_\mu\hat{f}_{1\beta}
+D\hat{f}_2
\right), \label{Pp}
\end{align}
and also the representation of $P_-$ on $\mathcal{H}_-$ is as follows:
\begin{align}
P_- = & 
\mathcal{D}^\mu\mathcal{D}_\mu\mathbb{I}_-
+D\hat{f}_0^\ast
-{\rm i}\sqrt{2}\bar{\lambda}_{\dot{\alpha}}\hat{\bar{f}}_1^{\dot{\alpha}}\nonumber \\
& +\sqrt{2}\bar{\theta}_{\dot{\alpha}}\left(
 \sqrt{2}\bar{\sigma}^{\mu\dot{\alpha}\alpha}((\mathcal{D}_\mu\lambda_\alpha)+\lambda_\alpha\mathcal{D}_\mu)\hat{f}_0^\ast
+{\rm i}\bar{\sigma}^{\mu\nu\dot{\alpha}}_{\ \ \ \ \dot{\beta}}F_{\mu\nu}\hat{\bar{f}}_1^{\dot{\beta}}
-{\rm i}\sqrt{2}\bar{\lambda}^{\dot{\alpha}}\hat{f}^\ast_2
\right) \nonumber \\
& + \ovl{\theta\theta}(-2\lambda\lambda\hat{f}^\ast_0
+\sqrt{2}\lambda^\alpha\sigma^\mu_{\alpha\dot{\alpha}}\mathcal{D}_\mu\hat{\bar{f}}_1^{\dot{\alpha}}
-D\hat{f}^\ast_2). \label{Pm}
\end{align}

The expansion of  $P_\pm$ in the form of Eq.(\ref{P}) is given by 
\begin{equation}
P_\pm = \mathbb{I}_\pm \partial^\mu\partial_\mu +\mathbb{A}_\pm^\mu\partial_\mu+\mathbb{B}_\pm. 
\end{equation}
Using the formulae (\ref{E})$\sim$(\ref{omega}), we obtain the following 
expressions:
\begin{align}
\mathbb{E}_+(x_+) = & \mathbb{B}_+-(\partial_\mu \omega_+^\mu)-\omega_{\mu}^+\omega^{\mu}_+ \nonumber\\
= & -D\hat{f}_0-{\rm i}\sqrt{2}\lambda^\alpha \hat{f}_{1\alpha}
+\sqrt{2}\theta^\alpha\left(
\frac{1}{\sqrt{2}}\sigma^\mu_{\alpha\dot{\alpha}}(\mathcal{D}_\mu\bar{\lambda}^{\dot{\alpha}})\hat{f}_0
+{\rm i}\sigma^{\mu\nu\ \beta}_{\ \ \alpha}F_{\mu\nu}\hat{f}_{1\beta}-{\rm i}\sqrt{2}\lambda_\alpha\hat{f}_2
\right) \nonumber\\
& +\theta\theta(-\frac{1}{\sqrt{2}}(\mathcal{D}_\mu \bar{\lambda}_{\dot{\alpha}})\bar{\sigma}^{\mu\dot{\alpha}\alpha}\hat{f}_{1\alpha}+D\hat{f}_2) \label{Ep},
\end{align}
\begin{align}
\mathbb{E}_-(x_-) = & \mathbb{B}_- - (\partial_\mu \omega^{\mu}_-) -\omega^-_\mu\omega^{\mu}_- \nonumber\\
= & D\hat{f}^\ast_0-{\rm i}\sqrt{2}\bar{\lambda}_{\dot{\alpha}} \hat{\bar{f}}_1^{\dot{\alpha}}
+\sqrt{2}\bar{\theta}_{\dot{\alpha}}\left(
\frac{1}{\sqrt{2}}\bar{\sigma}^{\mu\dot{\alpha}\alpha}(\mathcal{D}_\mu\lambda_\alpha)\hat{f}^\ast_0
+{\rm i}\bar{\sigma}^{\mu\nu\dot{\alpha}}_{\ \ \ \ \dot{\beta}}F_{\mu\nu}\hat{\bar{f}}_1^{\dot{\beta}}
-{\rm i}\sqrt{2}\bar{\lambda}^{\dot{\alpha}}\hat{f}^\ast_2
\right) \nonumber\\
& +\ovl{\theta\theta}(-\frac{1}{\sqrt{2}}(\mathcal{D}_\mu \lambda^\alpha)\sigma^\mu_{\alpha\dot{\alpha}}\hat{\bar{f}}_1^{\dot{\alpha}}
-D\hat{f}^\ast_2) \label{Em},
\end{align}
and the bundle curvature $\Omega^{\mu\nu}_\pm$ are expressed by
\begin{align}
\Omega^{\mu\nu}_+(x_+) = &
-{\rm i}F^{\mu\nu}\mathbb{I}_+ +\sqrt{2}\theta^\alpha\frac{1}{\sqrt{2}}(
\sigma^\nu_{\alpha\dot{\alpha}}(\mathcal{D}^\mu\bar{\lambda}^{\dot{\alpha}})
-\sigma^\mu_{\alpha\dot{\alpha}}(\mathcal{D}^\nu\bar{\lambda}^{\dot{\alpha}}
))\hat{f}_0 \nonumber\\
& +\theta\theta((\mathcal{D}^\mu\bar{\lambda}_{\dot{\alpha}})\bar{\sigma}^{\nu\dot{\alpha}\alpha}
-(\mathcal{D}^\nu\bar{\lambda}_{\dot{\alpha}})\bar{\sigma}^{\mu\dot{\alpha}\alpha}
)\hat{f}_{1\alpha},
\end{align}
\begin{align}
\Omega^{\mu\nu}_-(x_-) = & 
-{\rm i}F^{\mu\nu}\mathbb{I}_- +\sqrt{2}\bar{\theta}_{\dot{\alpha}}\frac{1}{\sqrt{2}}(
\bar{\sigma}^{\nu\dot{\alpha}\alpha}(\mathcal{D}^\mu\lambda_\alpha)
-\bar{\sigma}^{\mu\dot{\alpha}\alpha}(\mathcal{D}^\nu\lambda_\alpha
))\hat{f}_0^\ast \nonumber\\
& +\theta\theta((\mathcal{D}^\mu\lambda^\alpha)\sigma^\nu_{\alpha\dot{\alpha}}
-(\mathcal{D}^\nu\lambda^\alpha)\sigma^\mu_{\alpha\dot{\alpha}}
)\hat{\bar{f}}_1^{\dot{\alpha}}.
\end{align}
In the same way, we calculate $\mathbb{E}_\pm \mathbb{E}_\pm \Psi_\pm $, $\Omega^{\mu\nu}_\pm \Omega_{\mu\nu}^\pm\Psi_\pm$ and 
replace the components of $\Psi_\pm$ with extracting operators so that we
obtain the representations of $\mathbb{E}_\pm^2$, $\Omega_{\mu\nu}^\pm \Omega^{\mu\nu}_\pm$ on $\mathcal{H}_\pm$ as follows:
\begin{align}
\mathbb{E}_+^2(x_+) = & (D^2-{\rm i}\lambda^\alpha\sigma^\mu_{\alpha\dot{\alpha}}(\mathcal{D}_\mu\bar{\lambda}^{\dot{\alpha}}))\hat{f}_0
+(\cdots)^\beta\hat{f}_{1\beta}+(\cdots)\hat{f}_2 
+\sqrt{2}\theta^\alpha(\cdots)_\alpha\hat{f}_0 \nonumber \\
& +\sqrt{2}\theta^\alpha(
-{\rm i}\sigma^\mu_{\alpha\dot{\alpha}}(\mathcal{D}_\mu\bar{\lambda}^{\dot{\alpha}})\lambda^\beta
-\sigma^{\mu\nu\ \gamma}_{\ \ \alpha}\sigma^{\lambda\kappa\ \beta}_{\ \ \gamma}F_{\mu\nu}F_{\lambda\kappa}
+{\rm i}\lambda_\alpha(\mathcal{D}_\mu\bar{\lambda}_{\dot{\alpha}})\bar{\sigma}^{\mu\dot{\alpha}\beta}
)\hat{f}_{1\beta} 
+\sqrt{2}\theta^\alpha(\cdots)_\alpha \hat{f}_2 \nonumber\\
& + \theta\theta(\cdots)\hat{f}_0+\theta\theta(\cdots)^\alpha\hat{f}_{1\alpha}
+\theta\theta({\rm i}(\mathcal{D}_\mu\bar{\lambda}_{\dot{\alpha}})\bar{\sigma}^{\mu\dot{\alpha}\alpha}\lambda_\alpha+D^2)\hat{f}_2
\label{Ep2},
\end{align}
\begin{align}
\mathbb{E}_-^2(x_-) = &
(D^2-{\rm i}\bar{\lambda}_{\dot{\alpha}}\bar{\sigma}^{\mu\dot{\alpha}\alpha}(\mathcal{D}_\mu\lambda_\alpha))\hat{f}^\ast_0
+(\cdots)_{\dot{\beta}}\hat{\bar{f}}_1^{\dot{\beta}}+(\cdots)\hat{f}^\ast_2 
+\sqrt{2}\bar{\theta}_{\dot{\alpha}}(\cdots)^{\dot{\alpha}}\hat{f}^\ast_0 \nonumber \\
& +\sqrt{2}\bar{\theta}_{\dot{\alpha}}(
-{\rm i}\bar{\sigma}^{\mu\dot{\alpha}\alpha}(\mathcal{D}_\mu\lambda_\alpha)\bar{\lambda}_{\dot{\beta}}
-\bar{\sigma}^{\mu\nu\dot{\alpha}}_{\ \ \ \ \dot{\gamma}}\bar{\sigma}^{\lambda\kappa\dot{\gamma}}_{\ \ \ \ \dot{\beta}}
F_{\mu\nu}F_{\lambda\kappa}+{\rm i}\bar{\lambda}^{\dot{\alpha}}(\mathcal{D}_\mu\lambda^\alpha)\sigma^\mu_{\alpha\dot{\beta}}
)\hat{\bar{f}}_1^{\dot{\beta}} 
+\sqrt{2}\bar{\theta}_{\dot{\alpha}}(\cdots)^{\dot{\alpha}} \hat{f}^\ast_2 \nonumber\\
& + \ovl{\theta\theta}(\cdots)\hat{f}^\ast_0+\ovl{\theta\theta}(\cdots)_{\dot{\alpha}}\hat{\bar{f}}_1^{\dot{\alpha}}
+\ovl{\theta\theta}({\rm i}(\mathcal{D}_\mu\lambda^\alpha)\sigma^\mu_{\alpha\dot{\alpha}}\bar{\lambda}^{\dot{\alpha}}+D^2)\hat{f}^\ast_2
\label{Em2},
\end{align}
and 
\begin{align}
\Omega_+^{\mu\nu}\Omega_{\mu\nu}^+(x_+) = & -F^{\mu\nu}F_{\mu\nu}\mathbb{I}_+ +\sqrt{2}\theta^\alpha(\cdots)_\alpha \hat{f}_0
+\theta\theta(\cdots)^\alpha \hat{f}_{1\alpha}, \label{OpOp} \\ 
\Omega_-^{\mu\nu}\Omega_{\mu\nu}^-(x_-) = &
-F^{\mu\nu}F_{\mu\nu}\mathbb{I}_-
+\sqrt{2}\bar{\theta}_{\dot{\alpha}}(\cdots)^{\dot{\alpha}} \hat{f}^\ast_0
+\ovl{\theta\theta}(\cdots)_{\dot{\alpha}} \hat{\bar{f}}_1^{\dot{\alpha}}
   \label{OmOm}.
\end{align}

Since $\sigma_{\ \ \alpha}^{\mu\nu\alpha} = \bar{\sigma}^{\mu\nu\dot{\alpha}}_{\ \ \dot{\alpha}}=0$, 
we have from (\ref{Ep}) and (\ref{Em})
\begin{equation}
 {\rm Str}(\mathbb{E}_+) = -{\rm Tr}[D] + {\rm i}\sigma^{\mu\nu\alpha}_{\ \ \alpha}{\rm Tr}[F_{\mu\nu}] + Tr[D] = 0,\ \ 
{\rm Str}(\mathbb{E}_-) =0.
\end{equation}
As for the supertraces of $\mathbb{E}^2$, we have
\begin{align}
{\rm Str}(\mathbb{E}_+^2) = & {\rm Tr}
 \left((D^2-{\rm i}\lambda^\alpha\sigma^\mu_{\alpha\dot{\alpha}}(\mathcal{D}_\mu\bar{\lambda}^{\dot{\alpha}})) \right. \nonumber \\
& -(-{\rm i}\sigma^\mu_{\alpha\dot{\alpha}}(\mathcal{D}_\mu\bar{\lambda}^{\dot{\alpha}})\lambda^\alpha
-\sigma^{\mu\nu\ \gamma}_{\ \ \alpha}\sigma^{\lambda\kappa\ \alpha}_{\ \ \gamma}F_{\mu\nu}F_{\lambda\kappa}
+{\rm i}\lambda_\alpha(\mathcal{D}_\mu\bar{\lambda}_{\dot{\alpha}}\bar{\sigma}^{\mu\dot{\alpha}\alpha})\nonumber\\
& \left. +({\rm i}(\mathcal{D}_\mu\bar{\lambda}_{\dot{\alpha}})\bar{\sigma}^{\mu\dot{\alpha}\alpha}\lambda_\alpha+D^2)
\right) \nonumber \\
= & {\rm Tr}(2D^2-4{\rm i}\bar{\lambda}_{\dot{\alpha}}\bar{\sigma}^{\mu\dot{\alpha}\alpha}(\mathcal{D}_\mu \lambda_\alpha)
-F_{\mu\nu}F_{\lambda\kappa}
-\frac{\rm i}{2}\varepsilon^{\mu\nu\lambda\kappa}F_{\mu\nu}F_{\lambda\kappa}), \\
{\rm Str}(\mathbb{E}_-^2) = &
{\rm Tr}(2D^2-4{\rm i}\bar{\lambda}_{\dot{\alpha}}\bar{\sigma}^{\mu\dot{\alpha}\alpha}(\mathcal{D}_\mu \lambda_\alpha)
-F_{\mu\nu}F_{\lambda\kappa}
+\frac{\rm i}{2}\varepsilon^{\mu\nu\lambda\kappa}F_{\mu\nu}F_{\lambda\kappa}),
\end{align}
where we omit the surface terms and the difference between arguments $x_+$ and $x_-$ which cancel by the integration over 
the four-dimensional manifold $\int_M d^4 x$.  

The supertrace of $\Omega^{\mu\nu}\Omega_{\mu\nu}$ amounts to 
\begin{equation}
{\rm Str}(\Omega_\pm^{\mu\nu}\Omega_{\mu\nu}^\pm) =-{\rm Tr}[F^{\mu\nu}F_{\mu\nu}]{\rm Str}\mathbb{I}_\pm =0.
\end{equation}
The above supertraces perfectly coincide with those of the ref.\cite{paper1} so that the heat kernel coefficients 
and the super Yang-Mills action derived from them do so as well. 
The Seeley-DeWitt coefficients are given by 
\begin{equation}
a_0(P) = a_2(P) =0,
\end{equation}
and 
\begin{equation}
a_4(P) = \frac{1}{16\pi^2}\int_M d^4x {\rm Tr}(
2D^2-4{\rm i}\bar{\lambda}_{\dot{\alpha}}\bar{\sigma}^{\mu\dot{\alpha}\alpha}(\mathcal{D}_\mu\lambda_\alpha)
-F_{\mu\nu}F^{\mu\nu}).
\end{equation}
We rescale the vector supermultiplet as 
$\{A_\mu,\lambda_\alpha,D \}\rightarrow \{gA_\mu,g\lambda_\alpha,gD\}$ 
where $g$ 
is the gauge coupling constant 
and fix the constant $c_4 $ such that
\begin{equation}
\frac{c_4}{8\pi^2} =\frac{1}{g^2}.
\end{equation}
Finally, we obtain the following super Yang-Mills action: 
\begin{equation}
I_{\rm SYM} = \int_M d^4x
{\rm Tr}\left[
 -\frac{1}{2}F_{\mu\nu}F^{\mu\nu}- 2{\rm i}\bar{\lambda}_{\dot{\beta}}\bar{\sigma}^{\mu\dot{\beta}\beta}(\mathcal{D}_\mu\lambda_\beta) 
+ D^2
\right].
\end{equation}
\section{Conclusions}
In this paper, we have reconstructed the super Yang-Mills theory on NCG. 
We have reviewed our previous paper\cite{paper1} 
which formulated the theory by expressing chiral and antichiral superfields in the functional spaces 
$\mathcal{H}_\pm$ in the vector notation such as $(\varphi,\psi^\alpha,F)$ and re-expressed them in 
the superspace coordinate in (\ref{Psip}) and (\ref{Psim}). The elements in the algebra $\mathcal{A}_+$ 
and $\mathcal{A_-}$ which were previously represented by the matrix form are now expressed by the superfields 
in (\ref{ua}),(\ref{bua}). On the other hand, the representation of the whole functional space 
$\mathcal{H}_M=\mathcal{H}_+ \oplus \mathcal{H}_-$ 
and $\mathcal{H}_F$, which is the space of labels of left and right-handed 
matter particles, 
remain vector notation in (\ref{HM}). So the 
algebra $\mathcal{A}_M$, the Dirac operators $\mathcal{D}_M$ and $\mathcal{D}_F$  which act on the 
whole $\mathcal{H}_M$  and on $\mathcal{H}_F$ respectively are represented by matrices such as (\ref{DM}) and (\ref{DF}). 
However, Dirac operators $\mathcal{D}$ and $\bar{\mathcal{D}}$ which are the matrix components of $\mathcal{D}_M$ 
are now expressed by differentials of $\theta$ and $\bar{\theta}$ in (\ref{mD}) and (\ref{bmD}). 

Internally fluctuated Dirac operators are also given in the 
superspace coordinate by (\ref{tDM}) with (\ref{tildeD}),(\ref{tildebD}). The supersymmetric invariant products 
in $\mathcal{H}_+$ and 
$\mathcal{H}_-$ are represented in (\ref{suprop}) and (\ref{suprom}) in the superspace as well.  
These modified Dirac operators and supersymmetric invariant products give the kinetic 
terms of matter fields in (\ref{PLtDPL}),(\ref{PRtDPR}). 
In our definition of internal fluctuation, the finite space 
Dirac operator,i.e. mass matrices with respect to family index are not modified and mass terms of 
action are  derived by the supersymmetric invariant product in (\ref{Imass}). 

In order to represent the elliptic operator $P_\pm$ necessary to obtain their 
heat kernel expansion coefficients, we have introduced new operators 
which extract the components of chiral and antichiral superfields expressed by 
the superspace coordinate system. They are given in (\ref{fp02}), (\ref{fm02}). 
Using the extracting operators, supertraces of operators which act on $\mathcal{H}_\pm$ 
are also represented in the superspace. They are given in (\ref{Strp}), (\ref{Strm}). 
As, after calculation of $P_\pm \Psi_\pm$, we replace the components of $\Psi_\pm$ with 
the extracting operators, we can obtain the representation of $P_\pm$ in the superspace. 
 
On these preparations, we have calculated the supertrace of $\mathbb{E}$, $\mathbb{E}^2$, 
$\Omega^{\mu\nu}\Omega_{\mu\nu}$, and have obtained completely the same heat kernel 
expansion coefficients as ref.\cite{paper1}. 
So, we have also arrived at the same action of super Yang-Mills action as well. 

The methods that we have introduced in this paper, extracting oprators, 
the representations of elliptic operators and supertrace on $\mathcal{H}_\pm$ in the 
superspace coordinate system, will be applied straightforward to the other our supersymmetric models 
on NCG, i.e., minimal supersymmetric standard models and supergravity on 
NCG\cite{paper2,paper3}. 


\newpage 

\begin{thebibliography}{99}
\bibitem{connes1}
A. H. Chamseddine and A. Connes, Phys. Rev. Lett. {\bf 77}, 4868(1996),
\href{http://arxiv.org/pdf/hep--th/9606056.pdf}{[arXiv:hep-th/9606056]}.
\bibitem{connes2}
A. H. Chamseddine and A. Connes, Phys. Rev. Lett. {\bf 99}, 191601(2007),
\href{http://arxiv.org/pdf/0706.3690v3.pdf}{[arXiv:0706.3690]}.
\bibitem{connes4}
A. Connes, J. High Energy Phys. {\bf 0611},081(2006).
\bibitem{connes3}
A. H. Chamseddine and A. Connes, J. Geom. Phys. {\bf 58}, 38(2008), 
\href{http://arxiv.org/pdf/0706.3688v1.pdf}{[arXiv:0706.3688]}.
\bibitem{connes0}
A. Connes, {\it Noncommutative Geometry}, (Academic Press, New York, 1994).
\bibitem{connes7}
A. Connes, Comm. Math. Phys. {\bf 182}, 155(1996), 
\href{http://arxiv.org/pdf/hep-th/9603053v1.pdf}{[arXiv:hep-th/9603053]}.
\bibitem{connes10}
A. H. Chamseddine and A. Connes, J. Geom. Phys. {\bf 57}, 1(2006),
\href{http://arxiv.org/pdf/hep-th/0605011.pdf}{[arXiv:hep-th/0605011]}.
\bibitem{connes8}
A. H. Chamseddine and A. Connes, Comm. Math. Phys. {\bf 186}, 731(1997), 
\href{http://arxiv.org/pdf/hep-th/9606001v1.pdf}{[arXiv:hep-th/9606001]}.
\bibitem{martin}
S. P. Martin, \href{http://arxiv.org/pdf/hep-ph/9709356v6.pdf}{[arXiv:hep-ph/9709356]}.
\bibitem{paper0}
Hikaru Sato, S.Ishihara, H.Kataoka, A.Matsukawa and M.Shimojo, \\
Prog.Theor.Exp.Phys, \href{http://ptep.oxfordjournals.org/content/2014/5}{{\bf 053B02},(2014)}.
\bibitem{paper1}
Hikaru Sato, S.Ishihara, H.Kataoka, A.Matsukawa and M.Shimojo, 
Prog.Theor.Exp.Phys, \href{http://ptep.oxfordjournals.org/content/2014/7}{{\bf 073B05},(2014)}.
\bibitem{paper2}
M.Shimojo, S.Ishihara, H.Kataoka, A.Matsukawa and Hikaru Sato, 
Prog.Theor.Exp.Phys, \href{http://ptep.oxfordjournals.org/content/2015/1}{{\bf 013B01},(2015)}.
\bibitem{WessBagger}
J.Wess, J.Bagger, {\it Supersymmetry and Supergravity", Princeton University Press}, Princeton Series in 
Physics, (Princeton, New Jersey, 1992).
\bibitem{gracia}
J. Gracia-Bond\'ia, B.Iochum, and T.Shucker, Phys. Lett. {\bf B416}, 123(1998).
\bibitem{lizzi}
G. M. F. Lizzi, G. Mangano and G. Sparano, Phys. Rev. {\bf D55}, 6357(1997), 
\href{http://arxiv.org/pdf/hep-th/9610035.pdf}{[arXiv:hep-th/9610035]}.
\bibitem{gilkey}
P.Gilkey, {\it Invariance Theory, the Heat Equation and the Atiyah-Singer Index Theorem}, Mathematical Lecture Series 4, (Publish or Perish Press, Berkeley CA, 1984).
\bibitem{Lopez}
J.L.L\'{o}pez, O.Obreg\'{o}n, M.P.Ryan and M.Sabido, 
International Journal of Modern Phys. A {\bf Vol28}, Issue12(2013).
\bibitem{Fiorenzo}
Fiorenzo Bastianelli, (1991), \href{https://arxiv.org/abs/hep-th/9112035}{[arXiv:hep-th/9112035]}.
\bibitem{Berredo}
Guilherme de Berredo-Peixoto and Ilya L. Shapiro,
\href{http://arxiv.org/abs/hep-th/0307030}{[arXiv:hep-th/0307030]}.
\bibitem{paper3}
M.Shimojo, S.Ishihara, H.Kataoka, A.Matsukawa and Hikaru Sato, 
Prog.Theor.Exp.Phys, \href{https://academic.oup.com/ptep/article/3003364}{{\bf 023B02},(2017)}.
\end{thebibliography}
\end{document}